\documentclass[notitlepage, onecolumn, superscriptaddress, nofootinbib]{revtex4-1}
\usepackage{subfiles}
\usepackage{natbib}
\setcitestyle{numbers} 
\setcitestyle{square}
\usepackage{graphicx,amsmath,amssymb,subcaption}
\usepackage{color}
\usepackage{url}
\usepackage{booktabs}
\usepackage{float}
\usepackage{comment}
\usepackage[font=small,labelfont=bf,tableposition=top]{caption}
\DeclareCaptionLabelFormat{andtable}{#1~#2  \&  \tablename~\thetable}

\begin{document}	
	\title{Complexity of human response delay in intermittent control: The case of virtual stick balancing}
	\author{Takashi Suzuki}
	\email{d8181104@u-aizu.ac.jp}
	\affiliation{School of Computer Science and Engineering, University of Aizu, Japan}
	\author{Ihor Lubashevsky}
	\email{i-lubash@u-aizu.ac.jp}
	\affiliation{School of Computer Science and Engineering, University of Aizu, Japan}
	\author{Arkady Zgonnikov}
	\email{arkady@u-aizu.ac.jp}
	\affiliation{School of Computer Science and Engineering, University of Aizu, Japan}	
	\begin{abstract}
	Response delay is an inherent and essential part of human actions. In the context of human balance control, the response delay is traditionally modeled using the formalism of delay-differential equations, which adopts the approximation of fixed delay. However, experimental studies revealing substantial variability, adaptive anticipation, and non-stationary dynamics of response delay provide evidence against this approximation. In this paper, we call for development of principally new mathematical formalism describing human response delay. To support this, we present the experimental data from a simple virtual stick balancing task. Our results demonstrate that human response delay is a widely distributed random variable with complex properties, which can exhibit oscillatory and adaptive dynamics characterized by long-range correlations. Given this, we argue that the fixed-delay approximation ignores essential properties of human response, and conclude with possible directions for future developments of new mathematical notions describing human control.
	\end{abstract}
	
	\maketitle

\section{Introduction}
Human control often invokes discontinuous, intermittent mechanisms~\cite{cabrera2002onoff,loram2011human,gawthrop2011intermittent}. In car driving~\cite{lubashevsky2003bounded,markkula2018sustained}, postural sway~\cite{bottaro2005body,asai2009model}, balancing a stick on a fingertip~\cite{cabrera2002onoff,milton2009discontinuous}, etc., humans tend to hold off the control until the deviation of the controlled system from the desired state is large enough, and only then apply a corrective action. More continuous control strategies are less effective compared to such intermittent corrections, in part due to significant \textit{human response delays}~\cite{milton2008unstable,milton2016control}, which can reach hundreds of milliseconds. Indeed, throughout the control loop (stimulus perception, response execution), the speed of signal propagation in the nervous system is finite~\cite{nijhawan2008visual,campbell2007time}. In addition, responding to observed deviations has been shown to involve stochastic decision-making mechanisms~\cite{markkula2018sustained,zgonnikov2018evidence}, which require non-zero time to generate a response. For these reasons, humans cannot immediately respond to the input signal from the controlled system, and have to mitigate the effects of the resulting lag via control intermittency or other compensatory mechanisms~\cite{nijhawan2009compensating,desmurget2000forward,kawato1999internal,mehta2002forward}.

The crucial role of response delay in human intermittent control is now universally acknowledged: virtually all relevant models capture the delay effects in some way. By far the most popular approach adopts the formalism of delay differential equations (DDE) (see~\cite{myshkis1949general,erneux2009applied,lakshmanan2011dynamics} for general background, and~\cite{stepan2009delay,milton2011delayed} for discussion of response delay specifically in human balance control). DDE have a number of advantages compared to more sophisticated approaches (e.g., equations with distributed delay); most importantly, DDE can be efficiently solved numerically. At the same time, there is evidence that traditional DDE, albeit being useful in understanding human control, do not provide an accurate representation of human response delay.

The main drawback of the DDE-based models is that they effectively assume that human response delay is fixed, and cannot account for delay variability, which is observed experimentally in balancing literature~\cite{peterka2002sensorimotor,mehta2002forward}. Furthermore, even such experimental studies of response delay in human balance control tend to focus on \textit{average} delay values, putting little emphasis on delay variability. In a recent study of stick balancing on a fingertip, Milton et al.~\cite{milton2016control} estimated the delay to be 230 ms based on a task with a sensory blank-out, but the method used there required averaging across trials and therefore did not allow to estimate the delay variability. Even when full distributions of observed delays are reported, they tend to be ignored in the subsequent literature. For instance, in a seminal study of postural balance, Peterka mentioned that the average overall delay in the control loop was $206\pm11$ (SD) ms~\cite{peterka2002sensorimotor}. This figure (approx. 200 ms) was subsequently used as a point estimate in a number of postural control studies (e.g.,~\cite{asai2009model}), even though the actual values of the estimates reported in Ref.~\cite{peterka2002sensorimotor} were scattered in the range of approx. 100 to 250 ms depending on the subject and task parameters. Similar delays were found in another experimental paradigm, stick balancing. Mehta and Schaal~\cite{mehta2002forward} reported the average delays of 220 ms in real stick balancing, and 269 ms in a virtual version of the same task, but the inter-subject and inter-trial variability suggest that the actual delay values ranged from approx. 150 to 330 ms. Despite these findings, models of human intermittent control pervasively continue to employ point estimates of response delay and ignore its variability.

Besides substantial variability, research on response delay in human intermittent control is complicated by the notions that neural delays can possess non-stationary adaptation dynamics~\cite{eurich1999dynamics,eurich2002recurrent,eurich1999dynamics,eurich2000delay}, and that humans can employ predictive mechanisms to compensate for response delay~\cite{desmurget2000forward,kawato1999internal,mehta2002forward}. The original formalism of DDE cannot capture any of these features. One might argue that certain extensions of DDE can capture these properties individually, for instance, distributed delay~\cite{eurich2002recurrent,rahman2015dynamics}, stochastic time~\cite{ohira2007stochasticity}, time-varying delay~\cite{insperger2015semi, verriest2009stability}, predictive feedback~\cite{gawthrop2009predictive}. However, to mathematically describe all these properties together, one would necessarily need to concatenate multiple corresponding DDE extensions at the same time, which makes the hypothetical resulting model intractable. 

The overarching purpose of the present study is to call for development of principally new mathematical formalism describing human response delay in all its complexity. We support our case by the experimentally demonstrating how complex human response delay is even in a simple balancing task. Specifically, we directly measure response delay in a virtual stick balancing task previously used to study the mechanisms of human intermittent control~\cite{zgonnikov2014react}. Our results demonstrate, first, that the response delay is effectively a random variable which can deviate substantially from its mean value. Second, when the system dynamics is predictable, the delay can be negative, which suggests that the subjects sometimes anticipate the system behavior. Third, the response delay can exhibit oscillatory and adaptive dynamics characterized by long-range correlations. Based on the current results, we argue that development of new mathematical formalism is needed to accurately model human response delay in intermittent control. Such formalism needs to incorporate  as inherent properties 1) the stochastic nature of response delay, 2) its non-stationary and non-linear dynamics, and 3) possibility of prediction.

\section{Methods} \label{sec:methods}
We performed two experiments with two partially overlapping groups of subjects. The aim of Experiment 1 was to investigate the human response delay and its dynamics while controlling a system with unpredictable or partially predictable dynamics. The aim of Experiment 2 was to replicate the results of Experiment 1, and further investigate the delay dynamics in case the system dynamics is fully predictable.

The setup of the two experiments was identical, except for two modifications and an additional condition introduced in Experiment 2. In what follows, we first describe the general experimental design, and then underline the distinctions between the two experiments.

\subsection{Subjects}
Eight right-handed healthy male undergraduate students (age 20 to 21) participated in Experiment 1. Ten right-handed healthy male students (age 19 to 25) participated in Experiment 2. 
Experimental procedures were approved by the University of Aizu Ethics Committee. All subjects gave written informed consent to participate in the experiments.

\subsection{Stick balancing setup}
Virtual overdamped stick balancing was used a representative task involving human operator control over an unstable system. The experimental setup used here is based on the virtual balancing task previously used to study the control activation mechanisms~\cite{zgonnikov2014react, zgonnikov2015type}. Here we briefly outline the original stick balancing task and then describe in details how it was modified to measure human response delay in the current study.

\begin{figure}
	\centering
	\begin{subfigure}[t]{0.57\linewidth}
		\includegraphics[width=1\linewidth]{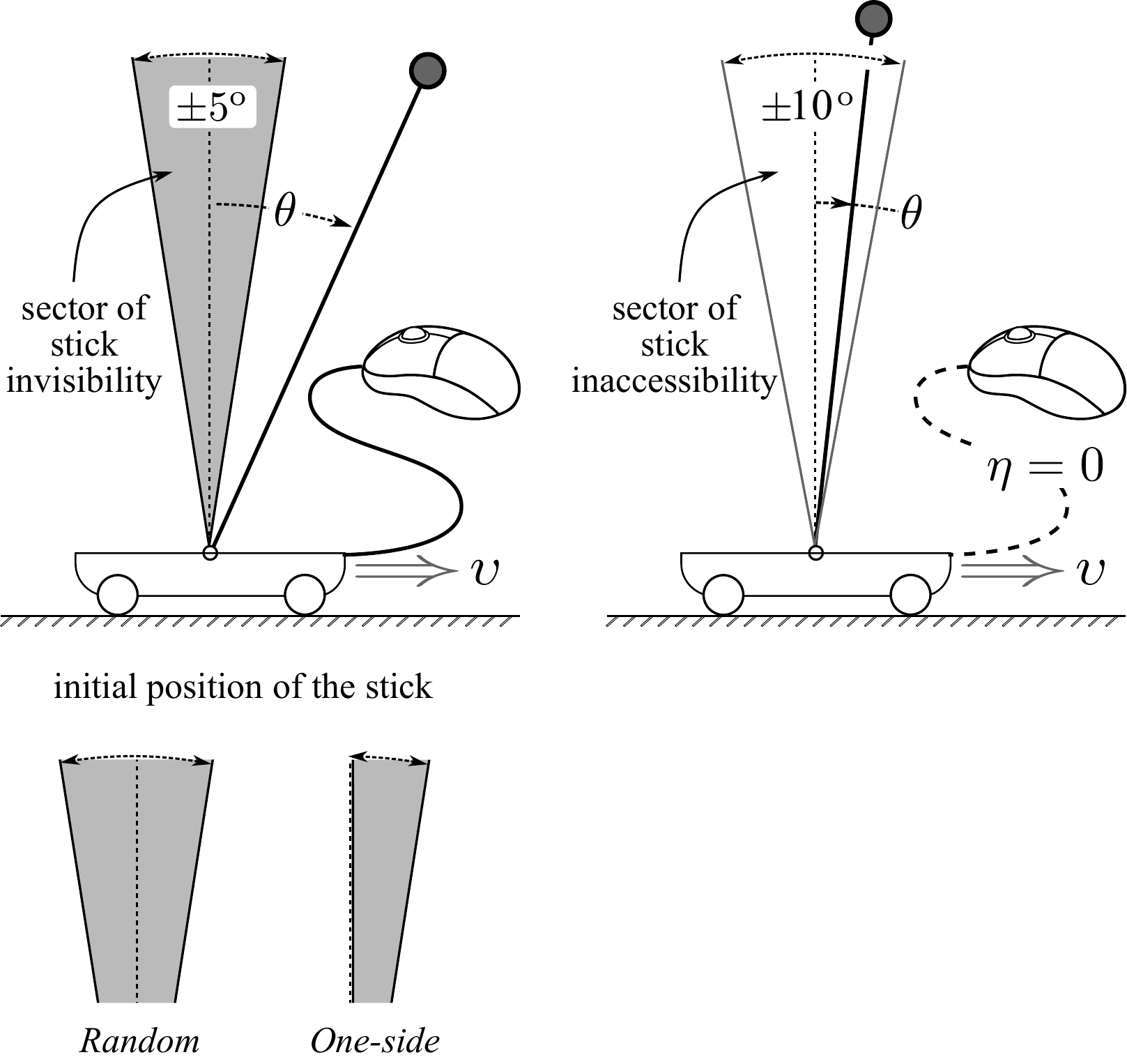}
		\caption{}
		\label{fig:stick}
	\end{subfigure}
	\hfill
	\begin{subfigure}[t]{0.37\linewidth}
		\includegraphics[width=1\linewidth]{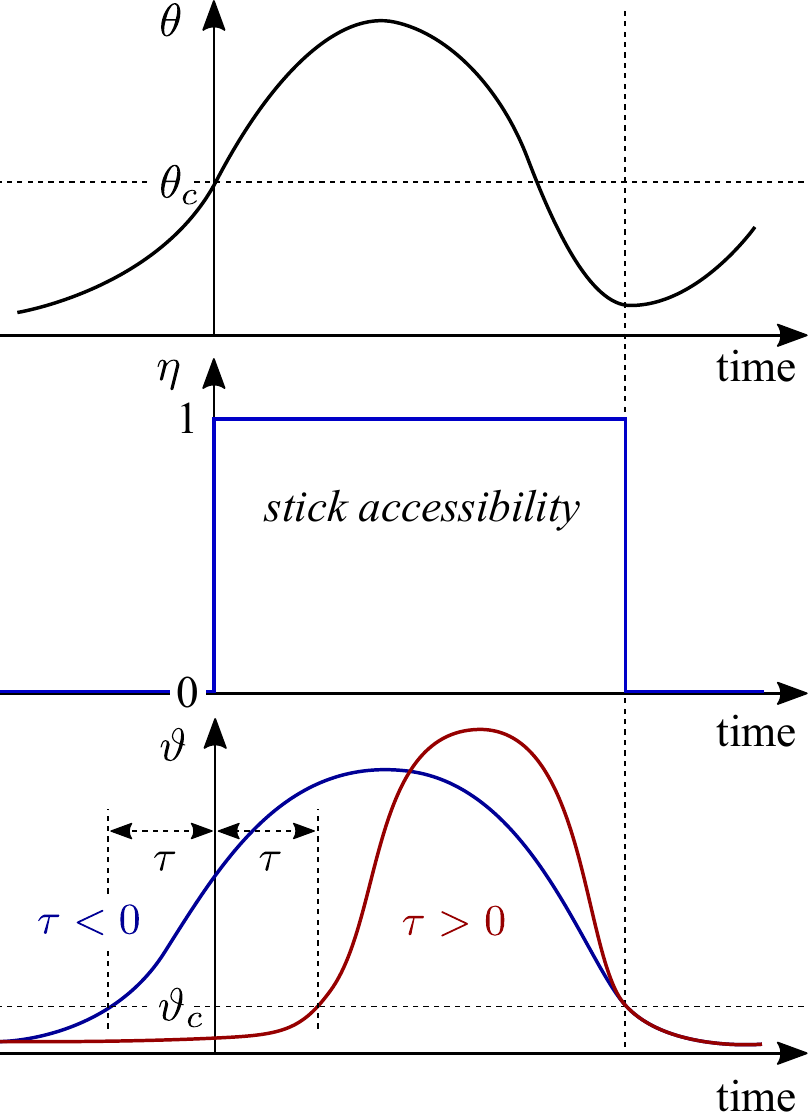}
		\caption{}
		\label{fig:delay}
	\end{subfigure}
	\caption{\textbf{(a)} Stick balancing task used in the random and one-side conditions(Experiments 1 and 2) and visible condition (Experiment 2). \textbf{(b)} Hypothetical trial dynamics in the visible condition (Experiment 2): stick angle $\theta$, stick accessibility $\eta$, and mouse velocity $v$.}  
	\label{fig:setup}
\end{figure}

On the computer screen, an operator observes a vertically oriented stick and a moving cart connected to the base of the stick via the pivot point (Fig.~\ref{fig:stick}). The task is to maintain the upright position of the stick by moving the cart horizontally via the computer mouse. The stick motion is simulated by numerically solving the equation (see Ref.~\cite{zgonnikov2014react} for details)
%
\begin{equation}
\tau_\theta \frac{d\theta}{dt} = \sin\theta -\frac{\tau_\theta}{l} v \cos\theta\,,
\label{eq:stick}
\end{equation}
where $\theta$ is the angle between the stick and its upright position, and $v$ is the cart velocity determined by the velocity of the mouse cursor controlled by the operator. The parameter $\tau_\theta$ characterizes the time scale of stick fall in the absence of subject's control. The stick length $l$ determines the characteristic magnitude of the cart's displacements required for keeping the stick upright. In the reported experiments we set $\tau_\theta = 0.5$~s, which corresponds to the typical time of stick fall without the operator's input about 2~s. The stick length was set to 60\% of the screen height (21.5 inch screen was used in the reported experiment).

\subsection{Response delay measurement}
\subsubsection{Experiment 1}
In Experiment 1, we investigated human response delay patterns under two conditions, random and one-side, each including 30 practice trials and 300 recorded trials. In both conditions, each trial started with the cart positioned in the middle of the screen and the stick placed inside the invisibility sector $|\theta_c| < 5^\circ$. The subjects' goal was to start controlling the stick as soon as it becomes visible, and return it to the vicinity of the vertical position (Fig.~\ref{fig:stick}). In the \textit{random} condition, the stick was initially deviated to the left or to the right of the vertical position, with the sign of deviation determined randomly prior to each trial ($\theta_0 = \pm 0.5^\circ$). In the \textit{one-side} condition, the stick was always deviated to the right of the vertical position ($\theta_0 = 0.5^\circ$). Therefore, in the random condition subjects could not predict where and when the stick would appear, but in the one-side condition the subjects knew the side but not the time of the stick appearance. Each trial continued for 5~s; in case the stick fell down, the trial was terminated earlier. Each trial was followed by a 3~s rest period. All subjects completed both random and one-side versions of the task, with the condition order counterbalanced (half the subjects completed the random condition first, and the other half completed the one-side condition first).

\subsubsection{Experiment 2}
Experiment 2 included the same random and one-side conditions as Experiment 1, with two modifications. First, when the stick was in the invisibility region, the mouse movements did not affect the cart position. This constraint was introduced to allow for registering predictive responses. Second, each trial was only terminated when the stick was returned to the invisibility sector ($|\theta|< \theta_c$) and, at the same, the cart velocity was sufficiently close to zero ($|v_n| < v_c$). This was done to ensure that the subjects respond to the detected stick deviation with a precise corrective movement, which would take into account not only the sign, but also the magnitude of the deviation.

In Experiment 2 we also introduced an additional condition, \textit{visible stick}. In this condition, the stick was visible at all times, but the subjects' mouse cursor movements did not affect the cart and the stick while the stick was in the inaccessibility sector $|\theta_c| < 10^\circ$ (Fig.~\ref{fig:stick}). In this way, the subjects could observe the stick motion and could potentially time their response to coincide with the moment the stick leaves the inaccessibility sector. 

As there was no evidence of condition order effects in Experiment 1, in Experiment 2 all subjects performed the experiment in the following order of conditions: visible, random, one-side. Each subject performed 30 practice and 400 recorded trials per condition.

To prevent small-scale, involuntary mouse movements to affect the stick motion in the invisibility/inaccessibility sector, we applied a threshold $v_c=0.01$ to mouse velocity during the experiment (the characteristic scale of mouse velocity fluctuations was about 0.5).

\subsection{Data analysis}
The response delay was defined as the time lag between the moment when the stick becomes visible/accessible and the moment when a subject starts to move the computer mouse, i.e., when the mouse velocity exceeds the cutoff value $v_c = 0.01$ (Fig.~\ref{fig:delay}). Note that even though in Experiment 2 the cart was inaccessible until the stick angle exceeded $\theta_c$, the subjects could still start moving the mouse in advance, trying to anticipate the stick appearance. In such cases, the delay would take negative values.

In Experiment 1, the subjects' mouse movements could affect the stick motion even when the stick was still in the invisibility region. For this reason, we excluded from our analysis the trials where the subjects moved the cart before the stick became visible, as these movements changed the controlled system dynamics, rendering the delay measurement impossible. The number of such trials varied across subjects (1\% to 45\%), and was on average 12\% in the random condition and 8\% in the one-side condition.

\section{Results}
\subsection*{Experiment 1}
We hypothesized that the random condition will result in larger response delays than the one-side condition. In the former, the direction of the stimulus is unknown, so the subjects' response presumably involved additional time needed to make a decision on the response direction. Indeed, in all eight subjects the response delay in the random condition was on average larger than in the one-side condition (Tab.~\ref{tab:exp_1}). 

\begin{table}[]
\centering
\caption{Mean delay values, standard deviations, and the results of the D'Agostino's K-squared normality test for delay times obtained in Experiment 1}
\label{tab:exp_1}
\begin{tabular}{@{}llllll@{}}
\toprule
Condition & Subject  & mean    & std      & $K^2$ statistic & $p$-value\\ \midrule
Random               &  1  &  0.29  &   0.08 &  53 &  3$\times10^{-12}$ \\
                     &  2  &  0.37   &   0.08 & 221 &  1$\times10^{-48}$ \\
                     &  3  &  0.29  &   0.12 &   9 &  9.44$\times10^{-4}$ \\
                     &  4  &  0.31  &   0.10 &  35 &  2.67$\times10^{-8}$ \\
                     &  5  &  0.41  &   0.22 &  30 &  2.76$\times10^{-7}$ \\
                     &  6  &  0.41  &   0.14 & 116 &  5.86$\times10^{-26}$ \\
                     &  7  &  0.39  &   0.13 & 121 &  4.91$\times10^{-27}$ \\
                     &  8  &  0.32  &   0.10 &  50 &  1.66$\times10^{-11}$ \\
One-side             &  1  &  0.27  &   0.09 &  13 &  1.91$\times10^{-3}$ \\
                     &  2  &  0.26    &   0.14 &  20 &  4.31$\times10^{-5}$ \\
                     &  3  &  0.24  &   0.13 &  14 &  8.11$\times10^{-4}$ \\
                     &  4  &  0.28  &   0.09 &  17 &  2.23$\times10^{-4}$ \\
                     &  5  &  0.26  &   0.14 &   9 &  9.32$\times10^{-4}$ \\
                     &  6  &  0.34   &   0.16 &  81 &  2.6$\times10^{-18}$ \\
                     &  7  &  0.33  &   0.11 &  58 &  2.93$\times10^{-13}$ \\
                     &  8  &  0.27  &   0.10 &  16 &  3.01$\times10^{-4}$ \\ \bottomrule
\end{tabular}
\end{table}

\begin{figure}
	\centering
	\begin{subfigure}[t]{0.44\linewidth}
		\centering
		\includegraphics[width=\textwidth]{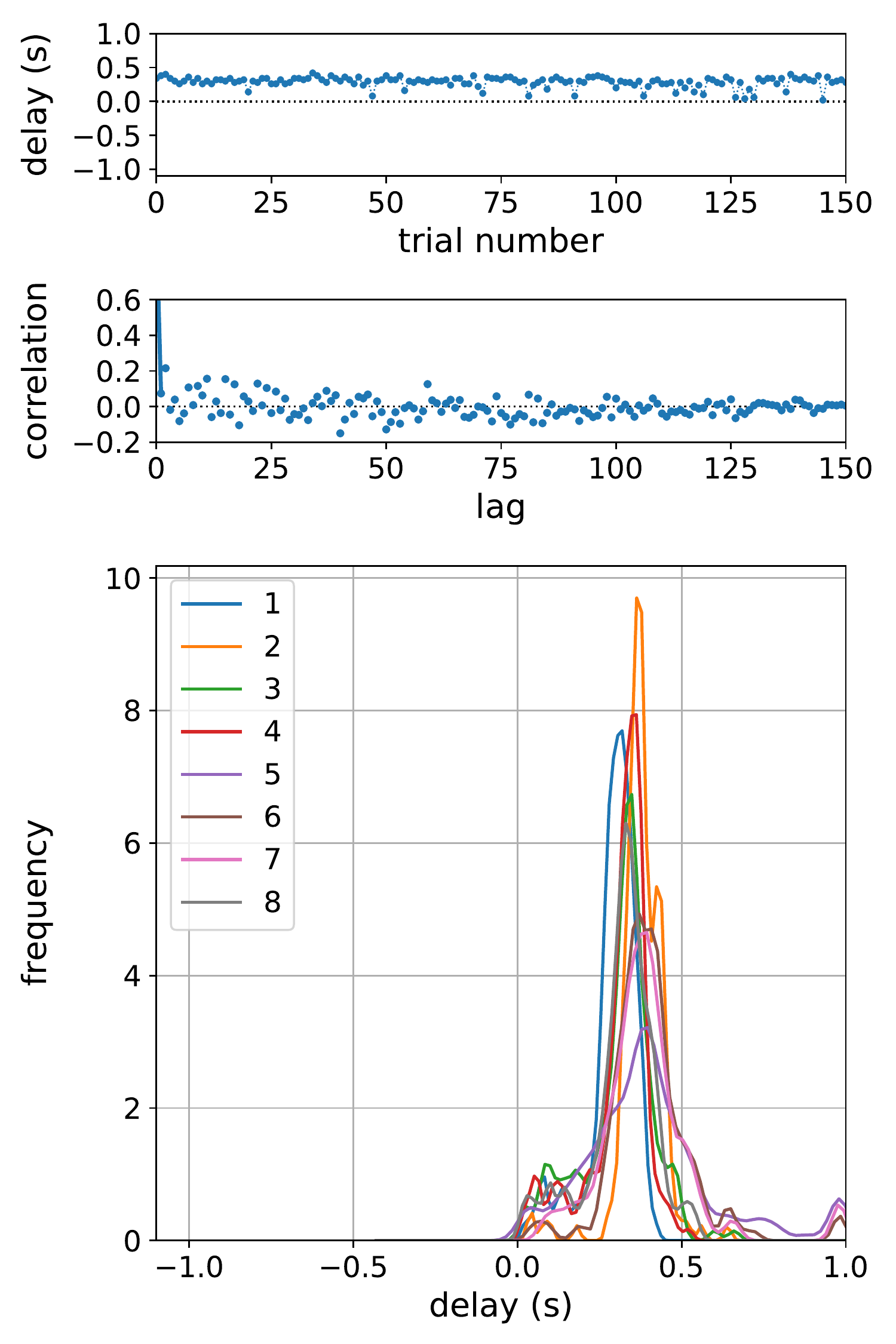}
		\caption{Random condition}
	\end{subfigure}
	~
	\begin{subfigure}[t]{0.44\linewidth}
		\centering
		\includegraphics[width=\textwidth]{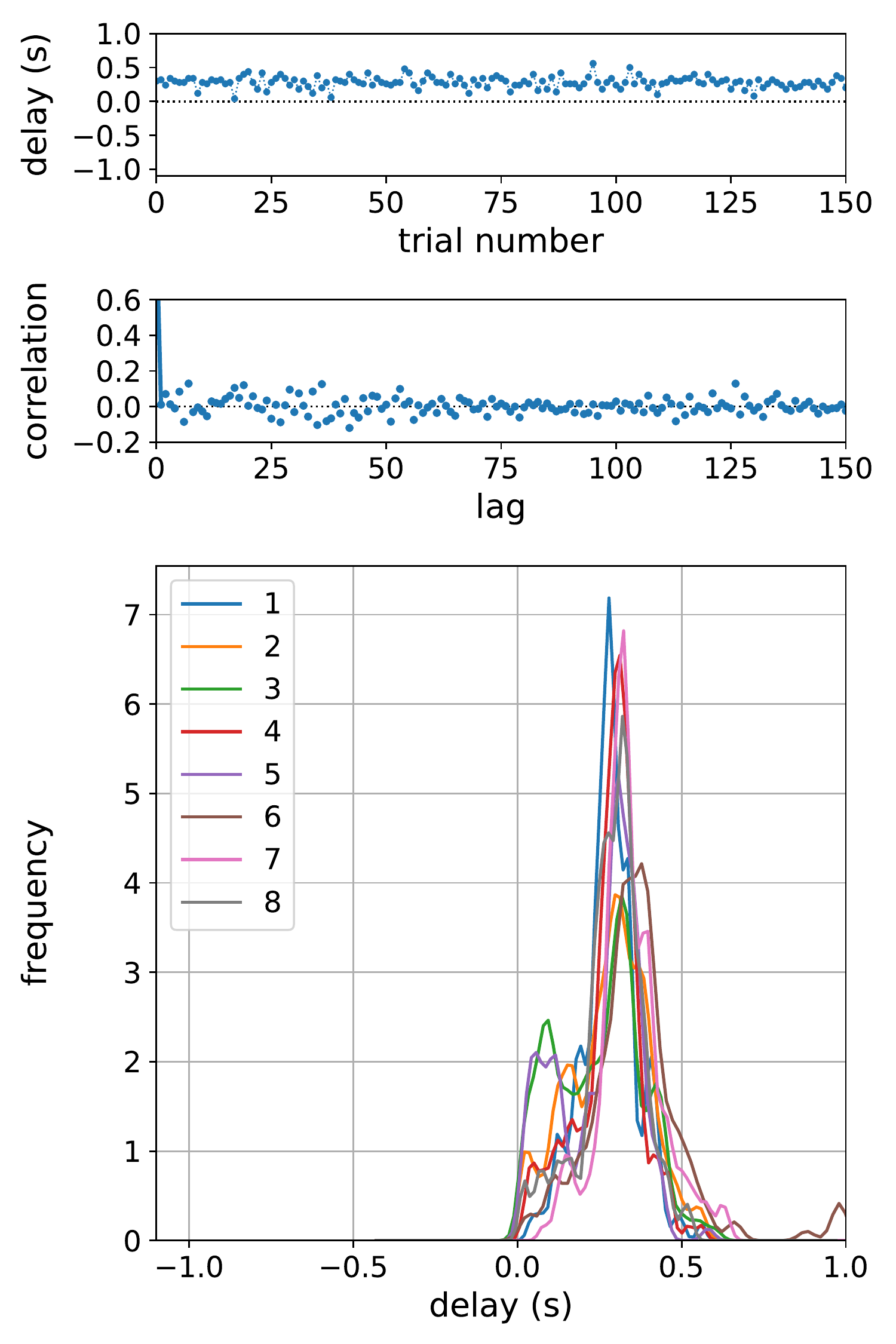}
		\caption{One-side condition}
	\end{subfigure}
	\caption{Results of Experiment 1: response delay dynamics, autocorrelation function, and distribution in (a) random and (b) one-side conditions.}
	\label{fig:results_exp1}
\end{figure}

Analysis of dynamics of the response delay across trials has shown that it exhibited roughly stationary behavior in both conditions in all subjects (Fig.~\ref{fig:results_exp1}, upper panels). The delay fluctuations were uncorrelated, showing little evidence for complex dynamics in Experiment 1 (Fig.~\ref{fig:results_exp1}, middle panels). 

\begin{table*}[]
\centering
\caption{Mean delay values, standard deviations, and the results of the D'Agostino's K-squared normality test for delay times obtained in Experiment 2}
\label{tab:exp_2}
\begin{tabular}{lllllll}
\hline
Condition & Delay dynamics type & Subject   & mean      & std      & $K^2$ statistic & p   \\ \hline
Random                &   Stationary          &   1    &   0.27    &   0.14    &   124.57    &   $8.91\times10^{-28}$ \\
                      &                       &   2    &   0.36    &   0.14    &   236.85    &   $3.70\times10^{-52}$ \\
                      &                       &   3    &   0.35    &   0.1    &   157.88    &   $5.20\times10^{-35}$ \\
                      &                       &   4    &   0.44    &   0.12    &   160.58    &   $1.35\times10^{-35}$ \\
                      &                       &   5    &   0.28    &   0.2    &   101.25    &   $1.03\times10^{-22}$ \\
                      &                       &   6    &   0.29    &   0.19    &   101.19    &   $1.06\times10^{-22}$ \\
                      &                       &   7    &   0.33    &   0.16    &   66.04    &   $4.56\times10^{-15}$ \\
                      &                       &   8    &   0.35    &   0.1    &   334.38    &   $2.46\times10^{-73}$ \\
                      &                       &   9    &   0.28    &   0.13    &   140.14    &   $3.71\times10^{-31}$ \\
                      &                       &   10    &   0.3    &   0.12    &   262.17    &   $1.18\times10^{-57}$ \\
One-side              &   Stationary          &   4    &   0.35    &   0.11    &   123.95    &   $1.21\times10^{-27}$ \\
                      &                       &   6    &   0.14    &   0.22    &   49.87    &   $1.48\times10^{-11}$ \\
                      &                       &   7    &   0.26    &   0.24    &   32.29    &   $9.72\times10^{-08}$ \\
                      &                       &   8    &   0.25    &   0.14    &   110.05    &   $1.26\times10^{-24}$ \\
                      &                       &   9    &   0.17    &   0.18    &   18.83    &   $8.13\times10^{-05}$ \\
                      &                       &   10    &   0.34    &   0.16    &   185.27    &   $5.88\times10^{-41}$ \\
                      & Oscillations          &   1    &   -0.44    &   0.39    &   29.17    &   $4.64\times10^{-07}$ \\
                      &                       &   2    &   0.02    &   0.26    &   12.73    &   $1.72\times10^{-03}$ \\
                      &                       &   3    &   0.0    &   0.19    &   12.44    &   $1.99\times10^{-03}$ \\
                      &                       &   5    &   -0.29    &   0.32    &   7.22    &   $2.71\times10^{-02}$ \\
Visible               &   Stationary          &   3    &   -0.25    &   0.15    &   69.86    &   $6.75\times10^{-16}$ \\
                      &                       &   5    &   -0.38    &   0.28    &   14.64    &   $6.63\times10^{-04}$ \\
                      &                       &   8    &   -0.05    &   0.11    &   140.28    &   $3.46\times10^{-31}$ \\
                      &                       &   9    &   -0.01    &   0.09    &   80.7    &   $2.99\times10^{-18}$ \\
                      &                       &   10    &   0.22    &   0.11    &   92.01    &   $1.05\times10^{-20}$ \\
                      & Oscillations          &  1    &   -0.32    &   0.36    &   2.24    &   $3.27\times10^{-01}$ \\
                      &                       &   4    &   -0.14    &   0.27    &   7.05    &   $2.95\times10^{-02}$ \\
                      &   Adaptation          &   2    &   -0.14    &   0.19    &   16.21    &   $3.02\times10^{-04}$ \\
                      &                       &   6    &   -0.22    &   0.26    &   10.87    &   $4.37\times10^{-03}$ \\
                      &                       &   7    &   0.02    &   0.19    &   39.66    &   $2.45\times10^{-09}$ \\ \hline
\end{tabular}
\end{table*}

\begin{figure}[t]
	\centering
	\includegraphics[width=0.44\linewidth]{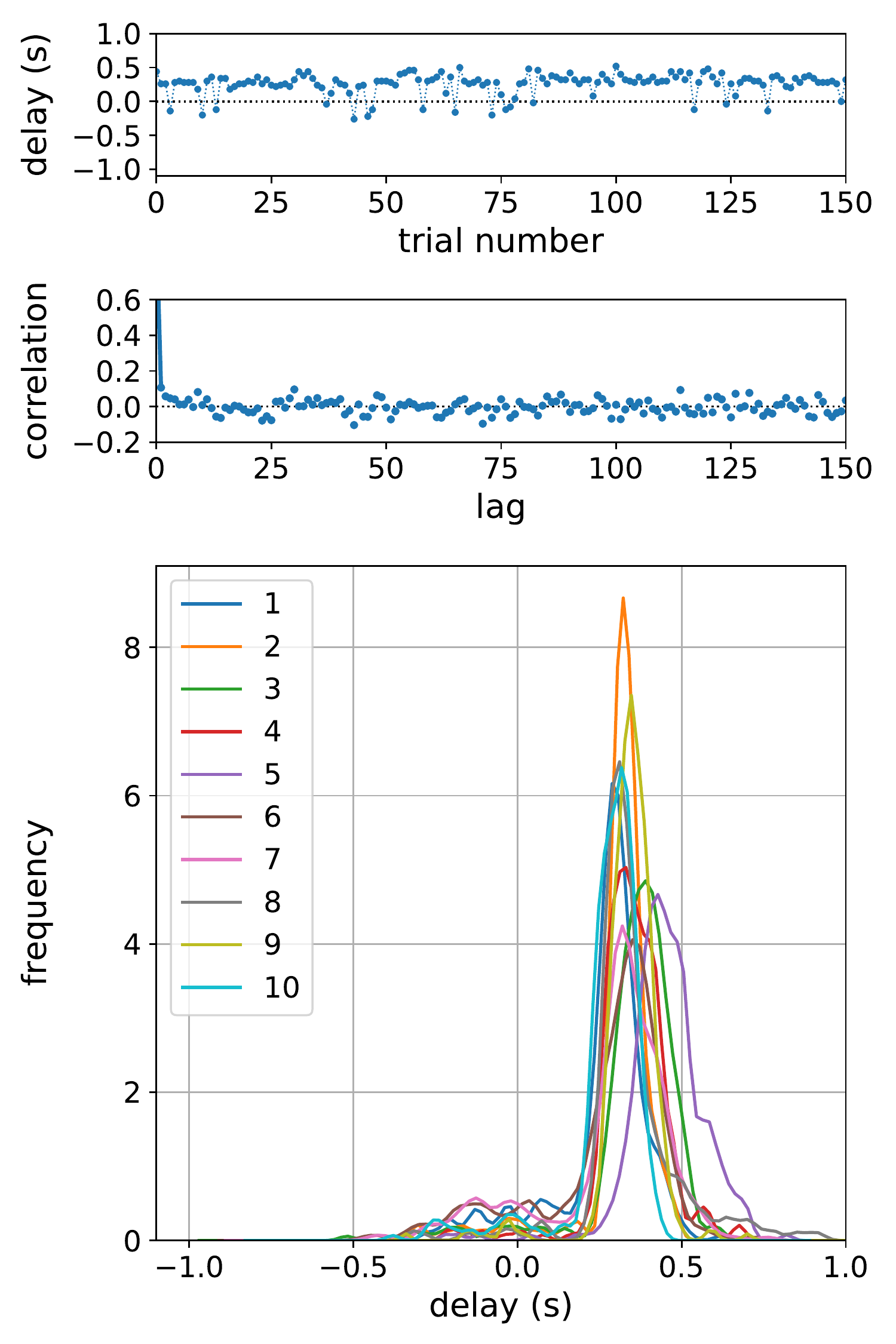}
	\caption{Results of Experiment 2 (random condition): response delay dynamics and autocorrelation function of a representative subject, and the histogram of delays produced by all subjects.}
	\label{fig:results_exp2_random}
\end{figure}

\begin{figure}[h]
	\begin{subfigure}[t]{0.44\linewidth}
		\centering
		\includegraphics[width=\textwidth]{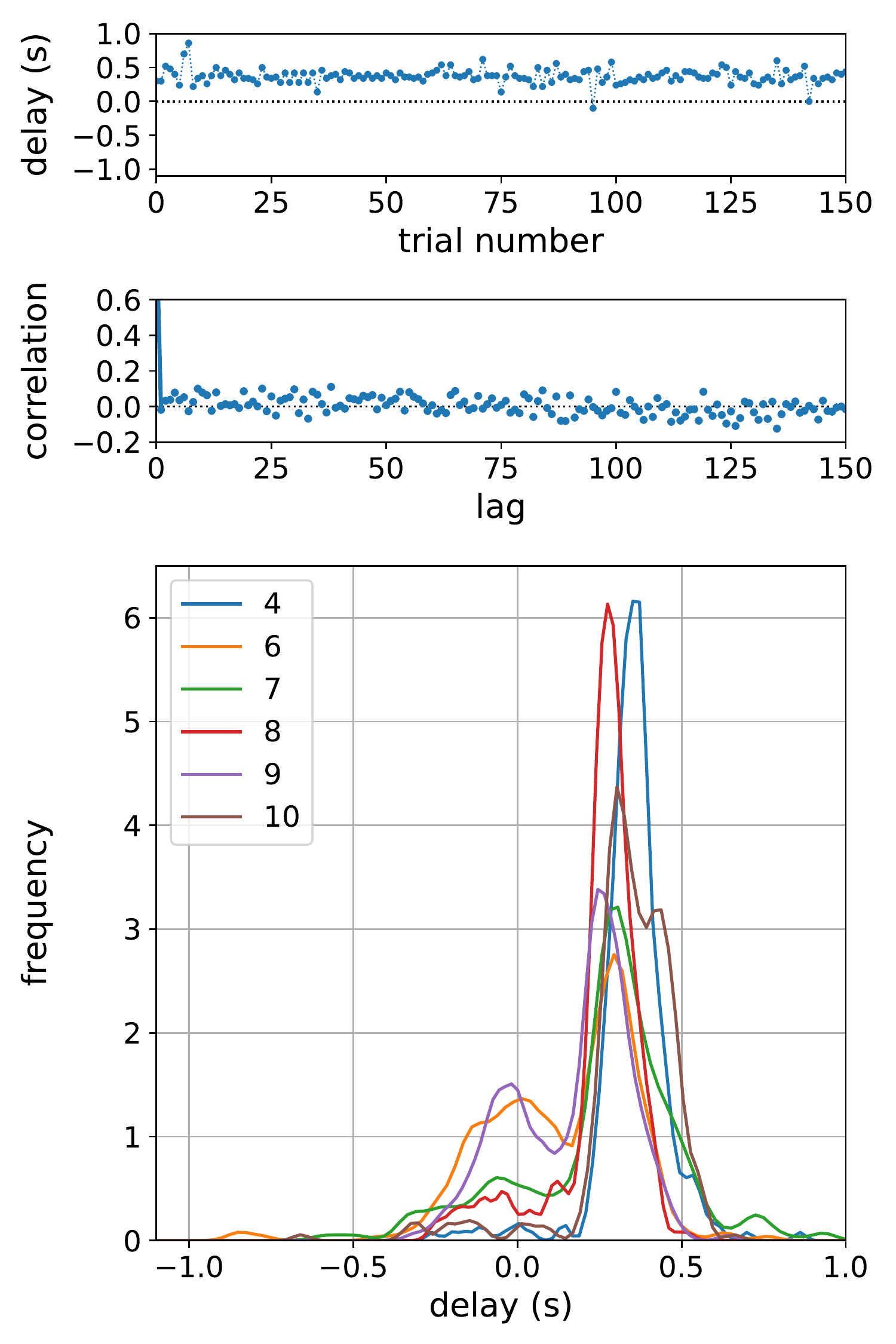}
		\caption{Stationary dynamics}
		\label{fig:results_exp2_oneside_st}
	\end{subfigure}
	~
	\begin{subfigure}[t]{0.44\linewidth}
		\centering
		\includegraphics[width=\textwidth]{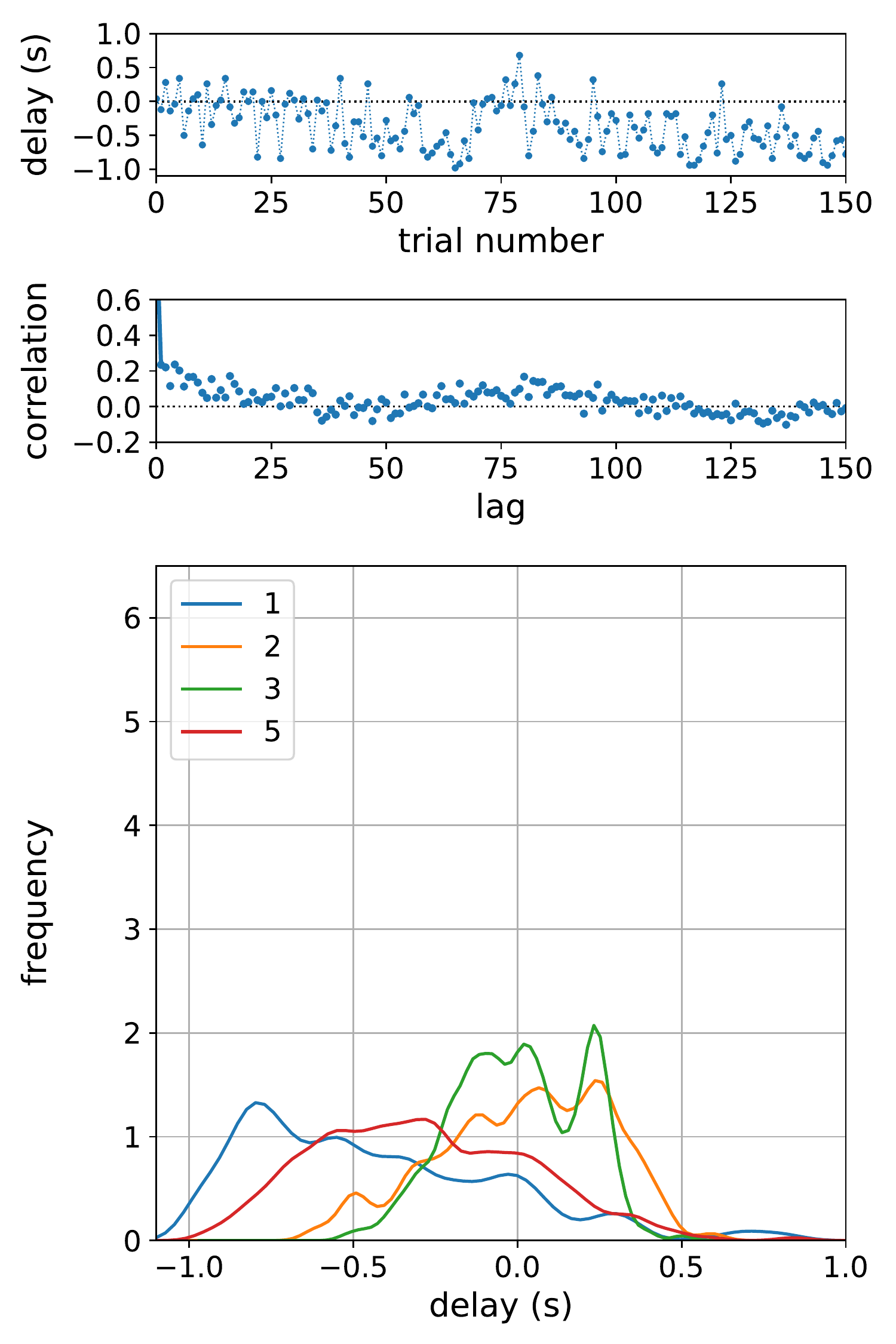}
		\caption{Oscillations}
		\label{fig:results_exp2_oneside_osc}
	\end{subfigure}
	\caption{Results of Experiment 2 (one-side condition): Response delay dynamics and autocorrelation functions of two representative subjects ((a) Subject 4 and (b) Subject 1), and the distributions of delays produced by all subjects. The subjects produced two types of delay dynamics: (a) stationary and (b) oscillatory.}
	\label{fig:results_exp2_oneside}
\end{figure}

The key finding of Experiment 1 is that the response delay exhibited substantial variability (Fig.~\ref{fig:results_exp1}, lower panels). In random condition, in most trials the delay ranged from approx. 200 ms to 500 ms. In the more predictable one-side condition, the delay ranged from 0 ms to 500 ms, indicating that the subjects often anticipated the moment when the stick leaves the invisibility region and utilized this to minimize the response delay. In both conditions, the within-subject standard deviation of the response delay was approx. $1/3$ of that subject's mean delay value. Finally, in all subjects and in both conditions, the D'Agostino $K^2$ test has shown significant evidence that the delay is not normally distributed (Tab.~\ref{tab:exp_1}). These findings provide evidence against the fixed delay approximation implied by the DDE-based models of intermittent control.

\subsection*{Experiment 2}
The purpose of Experiment 2 was two-fold: first, to replicate the results of Experiment 1 in a different subject group, and, second, to explore the properties of response delay in the situations where the dynamics of the controlled system is predictable. Experiment 2 thus involved an additional condition, \textit{visible} stick. In addition, in Experiment 2 we allowed for the subjects' response delay to be negative in the random and one-side conditions by disconnecting the mouse movements from the cart while the stick was in the invisibility sector. This enabled us to capture the signs of prediction mechanisms when the subjects controlled a partially predictable system.

The random condition results corroborate the findings of Experiment 1: In the random condition, the delay dynamics was largely stationary, with uncorrelated delay fluctuations (Fig.~\ref{fig:results_exp2_random}), and delay variability was characterized by standard deviations approx. $1/3$ to $1/2$ of the mean (Table~\ref{tab:exp_2}). Very few trials resulted in negative delay (cart movement initiated before the stick leaves the inaccessibility region), which suggests that predictions mechanisms were not efficient in the case of unpredictable stick motion direction.

The one-side condition results were also consistent with the corresponding results of Experiment 1, with two main distinctions. First, many subjects exploited the possibility of initiating the cart movement in anticipation of the stick appearance. In six subjects, this resulted in the response delay being negative in a substantial portion of trials (Fig.~\ref{fig:results_exp2_oneside}, lower panels). Second, we found that four subjects produced non-stationary time patterns of response delay. Particularly, the wave-like autocorrelation functions of the delay time series in Subjects 1, 2, 3, and 5 are the signs of periodic dynamics (Fig.~\ref{fig:results_exp2_oneside_osc}). One of the possible explanations for this is that these subjects regularly switched between two or more response strategies, e.g., responding to the stimulus only after it becomes visible vs. trying to predict the stimulus appearance in advance. 

\begin{figure}
	\centering
	\begin{subfigure}[t]{0.4\linewidth}
		\centering
		\includegraphics[width=\textwidth]{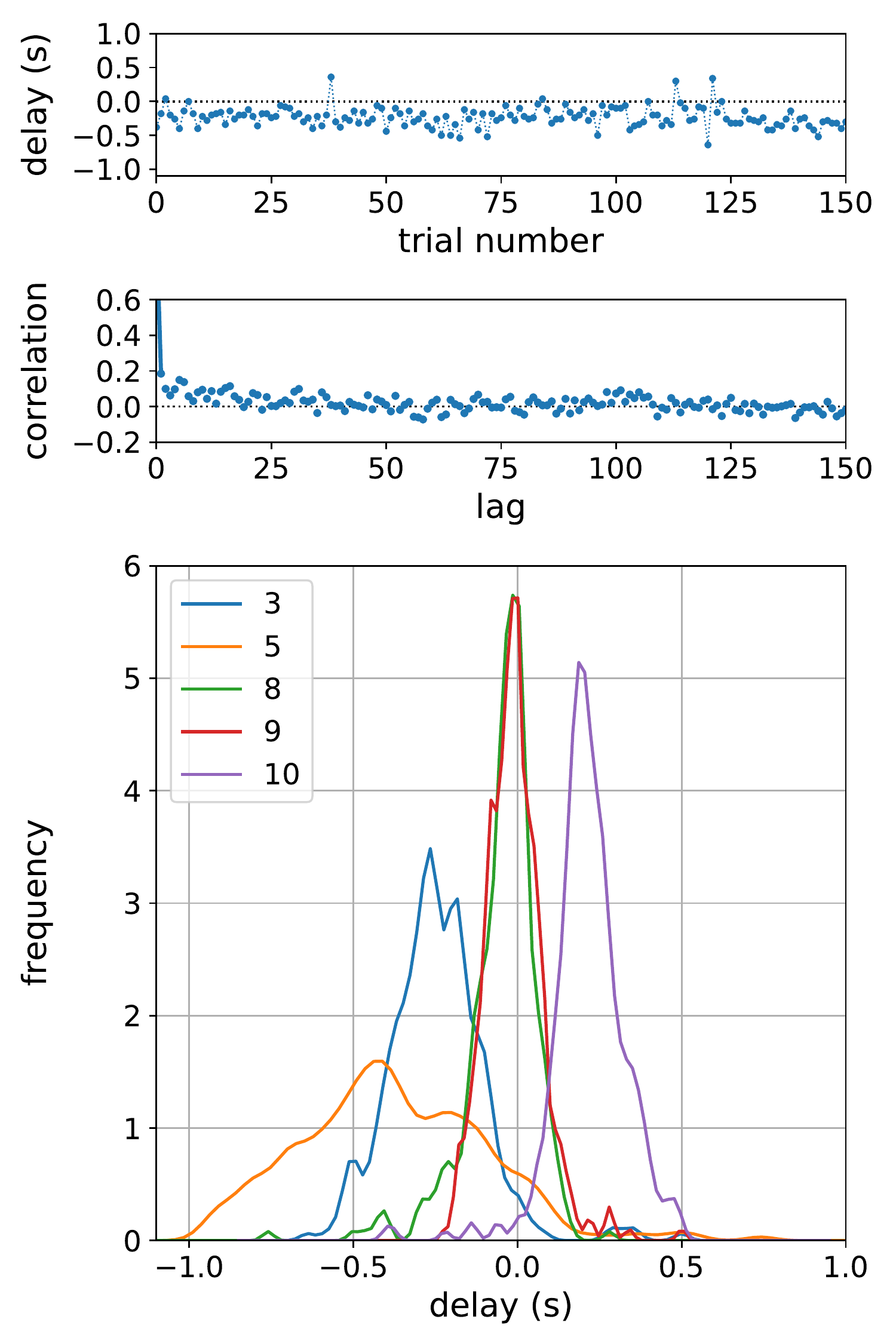}
		\caption{Stationary dynamics}
		\label{fig:results_exp2_visible_st}
	\end{subfigure}
	~
	\begin{subfigure}[t]{0.4\linewidth}
		\centering
		\includegraphics[width=\textwidth]{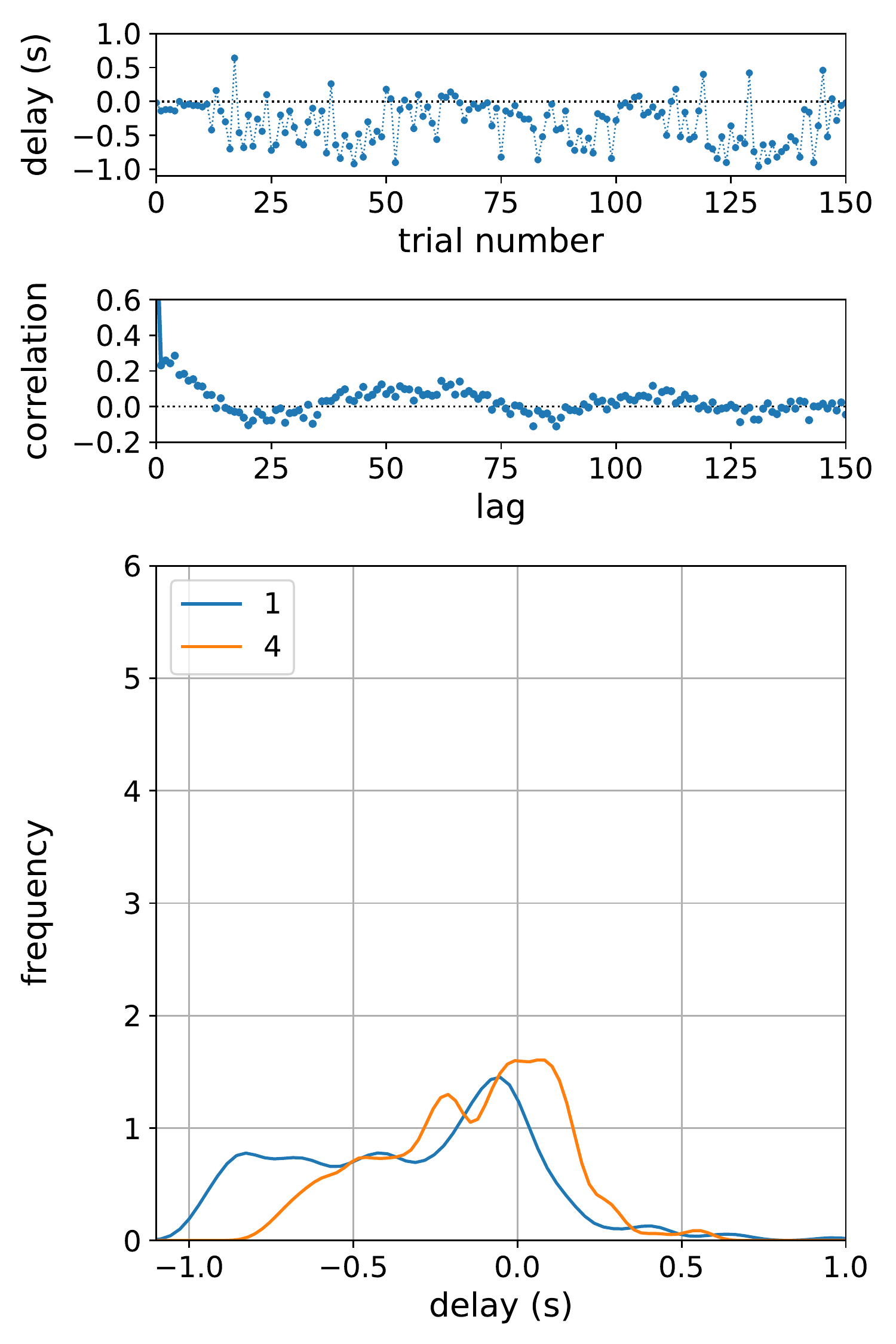}
		\caption{Oscillations}
		\label{fig:results_exp2_visible_osc}
	\end{subfigure}
	\\
	~
	\begin{subfigure}[t]{0.4\linewidth}
		\centering
		\includegraphics[width=\textwidth]{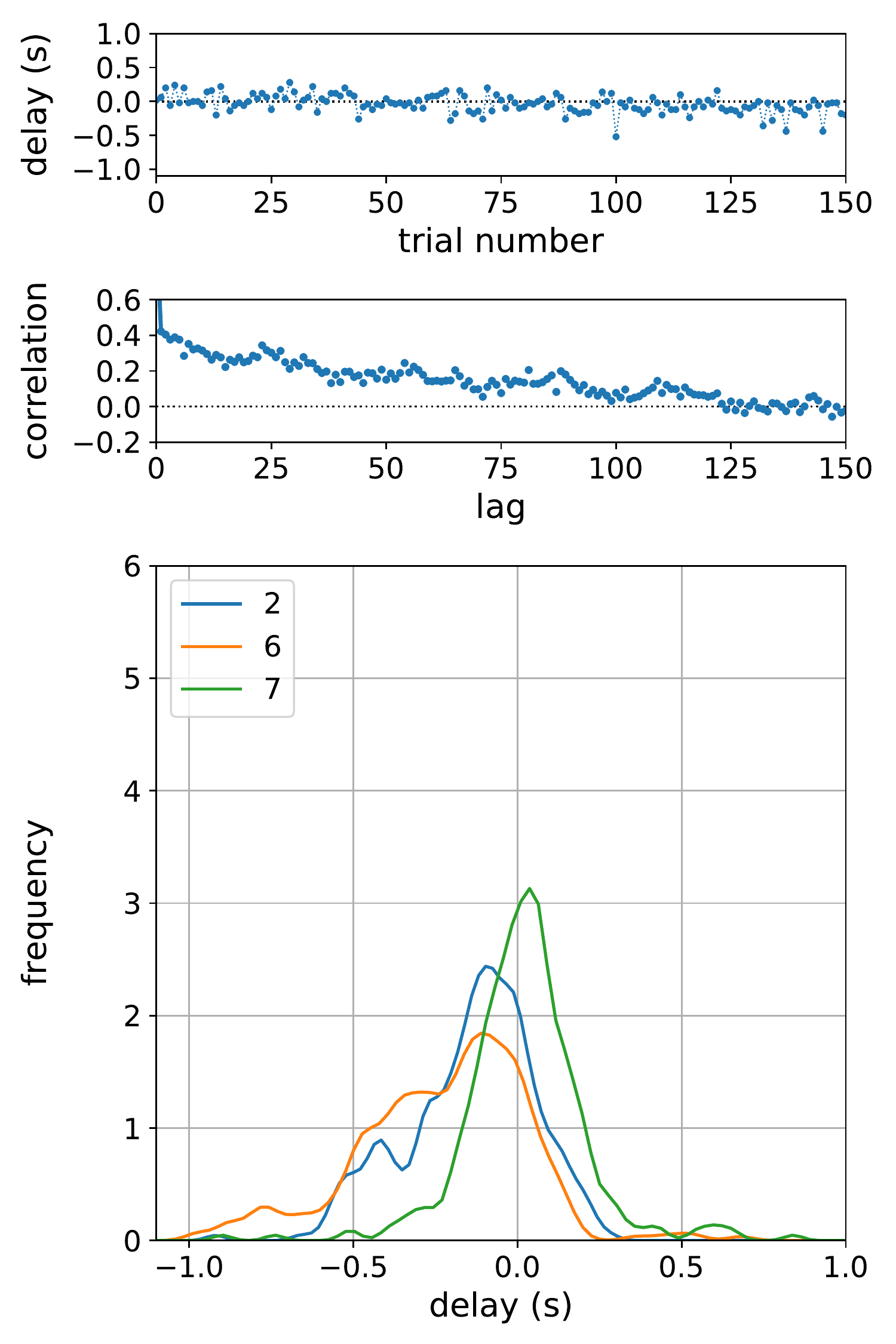}
		\caption{Adaptation}
		\label{fig:results_exp2_visible_ad}
	\end{subfigure}
	\caption{Results of Experiment 2 (visible condition): Response delay dynamics and autocorrelation functions of three representative subjects ((a) Subject 3, (b) Subject 1, and (c) Subject 2), and the distributions of delays produced by all subjects. The subjects produced three types of delay dynamics: (a) stationary, (b) oscillatory, and (c) adaptive.}
	\label{fig:results_exp2_visible}
\end{figure}

Finally, in the visible condition where the stick motion was fully observable and thus most predictable among all the conditions, we found an even more diverse spectrum of response delay dynamics. Similar to the one-side condition, some subjects employed the strategy of stationary response delay (Fig.~\ref{fig:results_exp2_visible_st}), and few subjects produced oscillating delay dynamics (Fig.~\ref{fig:results_exp2_visible_osc}). However, in addition to these two patterns observed in the one-side condition, we found that the response delay of three subjects showed the signs of adaptation. The time patterns of the delay produced by subjects 2, 6, and 7 show gradual decrease over time, which is characterized by long-range self-correlations. This suggests that these subjects learned the dynamics of the visible stick over time, and adapted their response correspondingly, which resulted in smaller delays.

\section{Discussion}
In this paper, we demonstrated that 
\begin{itemize}
	\item Human response delay is characterized by substantial individual differences;
	\item It can be distributed across the wide range of values;
	\item When the system under control is at least partially predictable, the delay can be compensated or over-compensated by anticipatory mechanisms;
	\item The delay can exhibit non-stationary and non-linear dynamics characterized by long-range correlations.
\end{itemize}
All these properties were observed together in a simple virtual balancing task. This suggests that in more sophisticated control processes, human response delay is at least as complex as in the task considered here. Studies of postural balance and stick balancing of a fingertip widely acknowledge the importance of response delay in human control, but almost universally utilize the fixed-delay approximation provided by delay-differential equations (e.g.,~\cite{cabrera2002onoff,bottaro2008bounded,asai2009model}). Our results imply that this approximation does not provide an adequate mathematical description of the human response delay.

Previously, uncertainty in human response timing was described in the model of random delay regarded in terms of random time with fixed delay or the delay changing randomly with time \cite{ohira2007stochasticity,1742-5468-2009-01-P01032}. The concept of time-varying delay, $\tau=\tau(t)$, is also met in describing the dynamics of a stem cell population~\cite{kirk1970cybernetics} and neural network behavior~\cite{wang2006synchronization} including random variations \cite[e.g.,][]{ASJC:ASJC1098}. There are also phenomena in population dynamics whose description turns to the idea of the delay time $\tau=\tau(x)$ depending on the current system state $x(t)$~\cite{ruan2006delay}. However, overall the models encompassing the idea of non-constant time delay are rare, possibly because such models are currently difficult to study both analytically and numerically. 

Intrinsic features of human sensorimotor system require the mathematical models describing human control to capture not only response delay, but also sensory deadzones (separately for position, velocity, and acceleration)~\citep{milton2013intermittent,Milton2015,gawthrop2009event}, noise~\cite{cabrera2002onoff,milton2008unstable}, and prediction~\citep{mehta2002forward,gawthrop2009predictive,gawthrop2011intermittent,Insperger2014}. Furthermore, such models may need to take into account the structure of the state space of the controlled system, e.g., saddle-type instability of the equilibrium in case of inverted pendulum with inertia~\cite{bottaro2008bounded,asai2009model,asai2013learning,suzuki2012intermittent,10.3389/fncom.2016.00034}. Capturing all these features together using the standard formalism of differential equations (stochastic and time-delayed) would inevitably result in a model which is mathematically intractable. For this reason, we argue that principally new mathematical formalism should be developed for describing human intermittent control and similar phenomena. 

Finally, we briefly outline a formalism that can potentially allow for the found features of human control in a natural way. The main drawback of the approaches to describing human control over unstable systems within the paradigm of delay differential equations is that this formalism does not take into account some of the basic features of human perception and cognition. Namely, the spatial point and the point-like instant of time are the main elements it deals with. However, we argue that these objects are ill-defined in the context of human cognition, so their use for describing human response might be problematic.

Indeed, first, our mental evaluation of a spatial position of a system under control with respect to its environment can be done only approximately; therefore, one may argue that in our consciousness instead of spatial points we operate with some regions with fuzzy boundaries. The concept of \textit{dynamical trap} --- a certain region where any point is treated by operator as equilibrium --- was developed as an alternative to the notion of stationary point to describe phenomenologically how humans maintain a system near its unstable equilibrium~\cite{lubashevsky2002long,Lubashevsky2003noise,lubashevsky2012dynamical}. It should be emphasized that the notion of dynamical trap is not reduced to the introduction of perception threshold because the control activation is an essentially probabilistic phenomenon~\cite{zgonnikov2014react,zgonnikov2015double,markkula2018sustained}. Even a simplified dynamical trap model \cite[e.g.,][]{lubashevsky2012dynamical} has to combine the elements of fuzzy control and the theory of stochastic processes to describe such human actions. Furthermore, dynamical traps can give rise to new type non-equilibrium phase transitions not meet in physical systems~ \cite{lubashevsky2002long,Lubashevsky2003noise,Lubashevsky2005chain,lubashevsky2016human}. The notion of dynamical trap is also the pivot point in the novel concept of noise-induced activation of human intermittent control~\cite{zgonnikov2014react, zgonnikov2015double}. 

Second, the analysis of  human temporal experiences also leads to the conclusion that our \textit{internal now} is not a point-like instant of time. The \textit{now} existing in the mind is represented by a finite time interval whose points are perceived as simultaneous --- the term \textit{specious present} is widely used for this human now. There is a vast literature devoted to such temporal experiences and a reader may be referred, e.g., to \cite{Phillips2017routage,sep-consciousness-temporal} for reviews and other references. For this reason the point-like instant should also replaced by a certain analogy to dynamical traps in time. 

The concept of space-time cloud \cite{ihor2017bookmind} --- a certain fuzzy region on the space-time continuum playing the role of \textit{here-and-now} in our consciousness and being an indivisible in it object --- allows for both these features. The real position of a given system and the current instant of time are hidden quantities for our mind and we are not able to operate with them directly. In this case the mental comparison of two objects can be implemented as the overlap of the two clouds.    

In this way we come to a conclusion that the desired formalism of describing response delays and uncertainty of spatial position should operate with space-time clouds and their dynamics. In this case all the effects discussed in the present work --- including the probabilistic nature of human response delay, its wide distribution, the contribution of immediate past and anticipation --- can potentially be allowed for directly. The spatial and temporal fuzziness of a space-time cloud may be represented in terms of some function $\Psi(x,v,t)$ whose dynamics is governed by the equations dealing with such functions as points of Hilbert space \cite[a preliminary discussion of this idea can be found in][]{lubashSuzuki2017cloud}. In this sense the formalism of space-time clouds resembles  that of quantum dynamics, however, it must be distinct from quantum dynamics in several aspects. One of them is fuzziness with respect to the temporal dimension caused by the finite thickness of human now. It reflects in that the persistence and change are experienced directly as mutually irreducible, complementary aspects of human perception \cite[e.g.,][]{Dainton2006stream}. For this reason, in contrast to quantum mechanics, the function $\Psi(x,v,t)$ contains the spatial coordinate $x$ together with the velocity $v$ in the list of its argument. Developing such models is the subject of future research.

	\textbf{All the data and program code produced in this study are available online:} \url{https://osf.io/eswf5/}.

	
\end{document}